\documentclass[12pt]{iopart}
\usepackage{graphicx}
\usepackage[authoryear,round]{natbib}
\usepackage{iopams}

\newcommand{\appropto}{\mathrel{\vcenter{
  \offinterlineskip\halign{\hfil$##$\cr
    \propto\cr\noalign{\kern2pt}\sim\cr\noalign{\kern-2pt}}}}}
    
\begin{document} 


\title[Model discrepancy by GP]{Efficient treatment of model discrepancy by
Gaussian Processes - Importance for imbalanced multiple constraint inversions}

\author{Thomas Wutzler$^1$}

\address{$^1$ Max Planck Institute for Biogeochemistry, 
Hans-Kn\"{o}ll-Stra\ss{}e 10, 07745 Jena, Germany}
\ead{twutz@bgc-jena.mpg.de}
\vspace{10pt}
\begin{indented}
\item[]December 2018
\end{indented}


\begin{abstract}
Mechanistic simulation models are inverted against observations in order to gain
inference on modeled processes. However, with the increasing ability to collect
high resolution observations, these observations represent more patterns of
detailed processes that are not part of a modeling purpose.
This mismatch results in model discrepancies, i.e.~systematic differences
between observations and model predictions. When discrepancies are not accounted
for properly, posterior uncertainty is underestimated. Furthermore parameters
are inferred so that model discrepancies appear with observation data stream
with few records instead of data streams correponding to the weak model parts.
This impedes the identification of weak process formulations that need to be
improved.
Therefore, we developed an efficient formulation to account for model
discrepancy by the statistical model of Gaussian processes (GP).
This paper presents a new Bayesian sampling scheme for model parameters and
discrepancies, explains the effects of its application on inference by a basic
example, and demonstrates applicability to a real world model-data integration
study.

The GP approach correctly identified model discrepancy in rich data streams.
Innovations in sampling allowed successful application to observation data
streams of several thousand records. Moreover, the proposed new formulation
could be combined with gradient-based optimization.
As a consequence, model inversion studies should acknowledge model
discrepancies, especially when using multiple imbalanced data streams. To this
end, studies can use the proposed GP approach to improve inference on model
parameters and modeled processes.
\end{abstract}

\vspace{2pc}

%
\submitto{\IP}
%
%

\section{Introduction}
The increased availability of high-resolution observation data streams in many
sciences \citep{Luo11a} allows constructing and evaluating detailed process
models \citep{Keenan11}. Often, the data supports higher detail than required by
the modeling purpose. Mismatch in detail between observations and model leads
to model discrepancy: systematic differences between the observed process and
the prediction of a calibrated model. Model discrepancy is often ignored in
model-data integration studies, or it is treated as part of a Gaussian residual
term \citep{Tarantola05}. Yet, model discrepancies cause correlation in
model-data residuals that are not accounted for by correlations in observations.
Such unaccounted correlations lead to overconfident estimates of parameters and
predictions \citep{Weston14}. They can also lead to biased parameter estimates,
especially when used with imbalenced data streams, i.e.~streams that strongly
differ by their number of records \citep{Wutzler14}. A treatment for model
discrepancy, hence, is important for model calibration. 

Researchers in numerical weather prediction early acknowledged the importance of
accounting for correlations in model data residuals and developed methods for
estimating those observations \citep{Bormann03, Healy05, Weston14}.
If a background, i.e.~a previous ensemble of model predictions, is available and
if certain conditions are fulfilled, the covariance matrix can be estimated
using both the predicions based on the background and predictions based on the
posterior sample \citep{Desroziers05, Waller14}.
However, it is hard for this approach to account for tradeoffs among imbalanced
data streams, because all ensemble members are based on the same estimate of the
covariance matrix. An approach that is suitable for trade-offs should allow
discrepancies and correlations matrices to differ across parameter
samples. A promising alternative approach is to represent discrepancy explicitly
as a statistical distribution called Gaussian process (GP) \citep{Kennedy01,
Brynjar14}, which is further described in section \ref{sec:methodsGP}.

Adding additional constraints or data streams of various types holds the promise
to constrain different aspects of the model \citep{Richardson10, Ahrens14a}. The
addition of sparse data stream, i.e.~data sets of relatively few records, has,
however, only neglibile influence on the model fit if no additional weights are
applied. Model discrepancy is allocated preferentially to sparse data streams.
This preferential allocation leads to complications in identifying model
deficiendies in studies using imbalanced data streams, i.e.~streams that
strongly differ in their number of records \citep{Wutzler14}.

The objectives of this study center around a better allocation of model
discrepancy and improved inference on modeled processes in inversion studies
using imbalanced data streams. The objectives are to
\begin{itemize}
\item demonstrate the problem of preferential
allocation of model discrepancy to sparse data streams,
\item present an efficient sampling formulation where
model discrepancy is represented as a Gaussian processes (GP),
\item demonstrate the ability of the GP approach to better allocate discrepancy
helping identification of which modelled processes need improvement.
\item demonstrate the applicability of the GP approach to real world problems
involving rich data streams.
\end{itemize} 

This study focuses on deterministic models and parameter estimation by batch
data assimilation \citep{Zobitz11}.
The initial state is assumed to be given or part of the vector of parameters to
estimate. For each proposed parameter vector, there is a unique model prediction
for each observation. 

In addition, the study focuses on approaches that combine evidence from all data
streams into a scalar valued objective function. The data stream likelihoods are
combined by their product, i.e.
saying how likely the first constraint ``AND'' the other constraints are, given
a set of parameters.
This approach differs from multi-objective optimization where the trade-offs are
explored by using a vector-valued objective function, and the user has to
specify additional information on how to select among different trade-offs
\citep{Miettinen99}.
The Bayesian approach employed in this study corresponds to using the
information content of the different streams to determine the location in the
pareto front in case of trade-offs.

The paper is structured as follows. The remainder of this section visually
introduces the approach of explicitly representing model discrepancy via a GP
and how this apoproach affects posterior predictions. 
Section \ref{sec:methods}
gives a mathematical introduction to the GP approach and introduces both the
basic example and the real-world ecosystem case used in the remainder of the
paper.
Section \ref{sec:resultsBasicExample} demonstrates the effects using the GP
approach on model inference with imbalanced data streams, by using a basic
example with two scenarios:
one that ignores model discrepancies and another one that explicitly models
discrepancy as a GP.
Section \ref{sec:resultsDalec} briefly demonstrates the applicability of the GP
approach to a real world inversion problem, the Dalec-Howland case.
Section \ref{sec:discussion} discusses how several problems were
tackled and what we learned.

\begin{figure}
\begin{center}\includegraphics{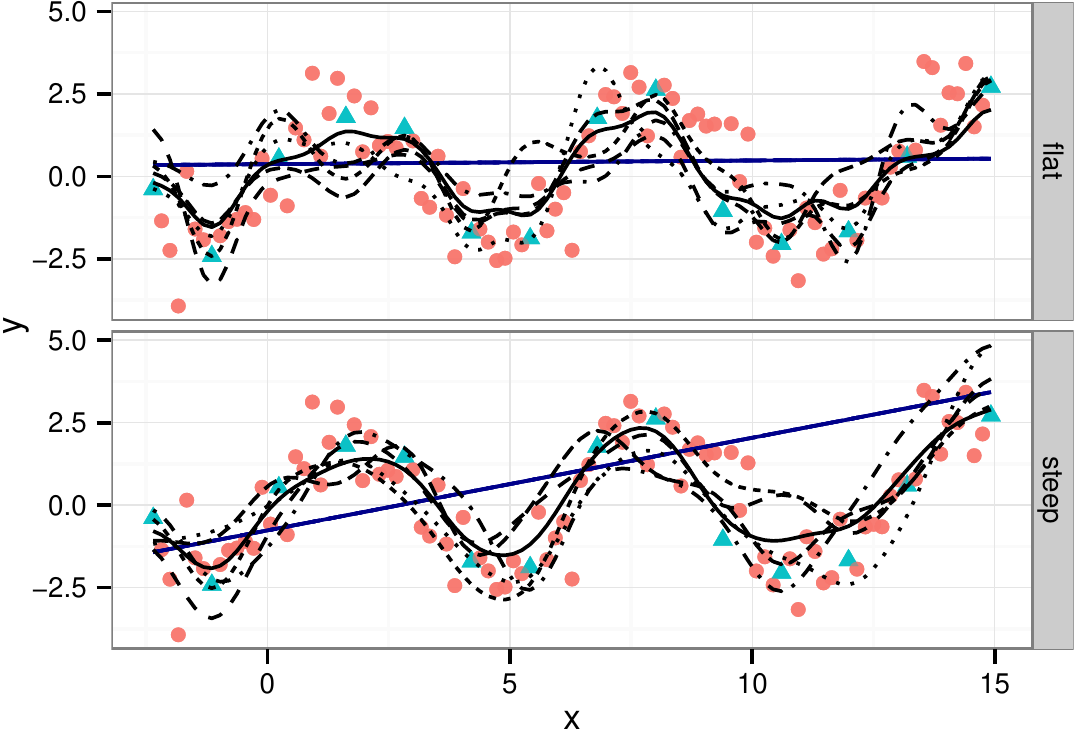}\end{center}
\caption{\label{fig:GPEx_randomFunction}
Process predictions (squiggly lies) that approximate observations (dots)
are the sum of a model prediction, $\mathbf{g}$ (straight line) and
several realizations of model discrepancy $\bdelta$: $\mathbf{o} \sim N(
\mathbf{g}(\btheta) + \bdelta, \sigma^2_{\epsilon})$.
Process predictions (top and bottom panel) obtained by different
parameters (model parameters $\btheta=(intersept, slope)^T$, and correlation length
$\psi$) are more similar than the corresponding model predictions. 
The respective two Gaussian process (GP) models of discrepancies have been
trained on a subset of data at supporting locations indicated by triangles.
}
\end{figure}

The concept of representing model discrepancy as a GP is visualized in Figure
\ref{fig:GPEx_randomFunction}, which displays two examples of model predictions
and model discrepancies against observations. The surrogate process of the
example, i.e.~the process generating the synthethic data, consists of a linear
part plus an oscillating part. The model only accounts for the linear part, but
is used to gain inferences on the slope. Model discrepancy constitutes the other
oscillating part. It is treated as a smooth sequence of values at observation
locations, $x$, and is represented as a GP. It is extremely large, here, for
visualization purposes. Its uncertainty is reflected by several samples from the
GP (several squiggly lines in Figure \ref{fig:GPEx_randomFunction}) conditioned on
observed model-observation differences at some supporting locations (triangles
in Figure \ref{fig:GPEx_randomFunction}). Note that differences between
observation and process predictions, i.e.~model plus discrepancy, are similar
across predictions by different model parameters (top and bottom panels).

\begin{figure} 
\begin{center}\includegraphics{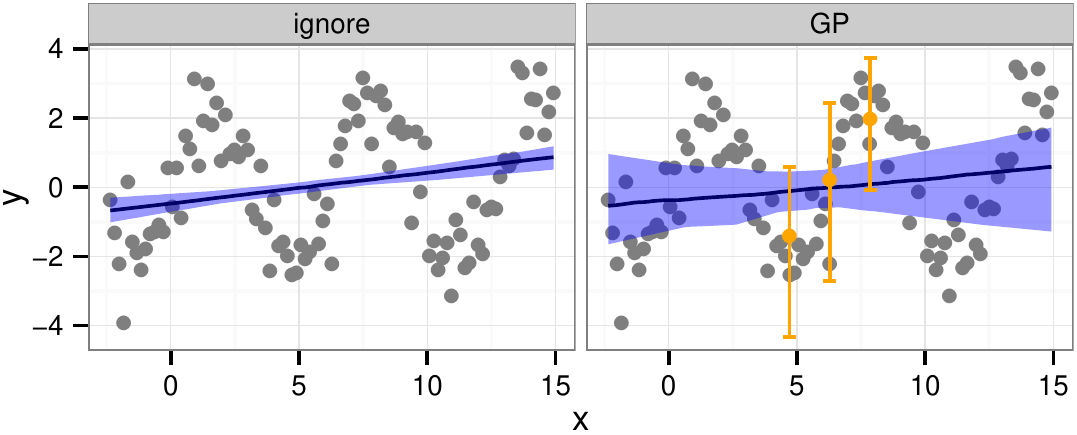}\end{center}
\caption{\label{fig:GPEx_posterior} Uncertainty estimates of
predictions change
when accounting for correlations due to model discrepancy. 
Sampling the linear part of the example data (dots, same as in Figure
\ref{fig:GPEx_randomFunction}) underestimated
prediction uncertainty when ignoring model discrepancy (left), as denoted by the
narrow shaded 95\% confidence band.
Representing discrepancy as a Gaussian process (GP) (right) increased the
estimate of uncertainty. In addition, the representation allowed inference on
process predictions ($\mathbf{g}_i(\btheta) + \bdelta_i$), as denoted by the
95\% confidence interval whiskers.
}
\end{figure}

The GP approach affects inference mainly by an increased estimate of prediction
uncertainty (Figure \ref{fig:GPEx_posterior}). The inversion with the ignore
scenario, which does not acknowledge correlation among model-data residuals,
overestimates information content in the observations and hence overestimates
precision of posterior estimates. Contrary, the GP approach strongly reduces
correlations among residuals between observations process predictions,
i.e.~model predictions plus discrepancies. At the same time, it yields in
a similar likelihood for a broader range of model parameters (compare rows in
Figure \ref{fig:GPEx_randomFunction}), and hence increases the estimate of
uncertainty.

\section{Methods\label{sec:methods}}
\subsection{modeling discrepancy as a Gaussian process (GP)
\label{sec:methodsGP}}

The vector of observations $\mathbf{o}$ of one data stream is modelled as
the sum of the deterministic model prediction $\mathbf{g}$, an unknown smooth
correlated vector of model discrepancy $\bdelta$, and observation error
$\bepsilon$ as \citep{Kennedy01}:
\numparts
\begin{eqnarray}
\label{eq:misfit}
\mathbf{o} &= \mathbf{g}(\boldsymbol{\btheta}) + \bdelta +
\bepsilon
\\
\bepsilon &\sim N(\mathbf{0}, \bSigma )
\\
\bdelta &\sim GP \left(\mathbf{0}, K \right)
.
\end{eqnarray}
\endnumparts

The model prediction $\mathbf{g}$ depends on parameter vector
$\boldsymbol{\btheta}$, whose distribution is to be estimated.
The model can be a simple regression function or a complex deterministic
dynamical simulation model.
The observation error $\bepsilon$ is assumed to be Gaussian noise, often with
the additional assumptions of independent errors: $\bSigma = \sigma^2_\epsilon
\mathbf{I}$.
The unknown vector of the model discrepancy is modelled as as GP. It accounts
for the additional correlations in model-data residuals due to model
discrepancy. Here, the GP uses a squared exponential covariance function,
$K(x_p,x_q)=\sigma^2_d exp(-(x_p-x_q)^2/\psi^2 + \sigma^2_\epsilon
\delta_{pq})$, where the Kronecker delta $\delta_{pg}$ equals
one for $p=q$ and zero otherwise. The covariance function has two
hyperparameters:
correlation length $\psi$ that controls the smoothness of the model discrepancy
across neighboring locations, and signal variance $\sigma^2_d$ that controls the
amplitude of the discrepancies varying around the expected value of zero
\citep{Rasmussen06}. The approach is applicable also for different assumptions
about the measurement error or the covariance structure of the model
discrepancies.

\subsection{Dimensionless model discrepancy variance
\label{sec:dimensionlessSigma}}

When dealing with multiple data streams, the different units of model
discrepancy variance $\sigma^2_d$ of different streams hinder the comparison of
model error. For example one stream may report weights of soil carbon and the
other fluxes of carbon dioxide per day.
Here, we propose normalizing discrepancy variance for stream, $k$, by the mean
variance of observation uncertainty,
\begin{equation}
\label{eq:sigmaK}
\sigma^2_{k} = \sigma^2_{d,k} / \sigma^2_{\epsilon,k}
. 
\end{equation}
The unitless quantity $\sigma^2_{k}$ allows comparing model discrepancy across
streams. For example, one can state that model discrepancy in stream $k_1$ is double the
model discrepancy of stream $k_2$, although they report different qualities.

\subsection{Parameters sampling 
\label{sec:parameterSampling}}

The parameters to estimate comprise the vector of model parameters,
$\boldsymbol{\btheta}$, and for each data stream, $k$, the two
scalar hyperparameters $\sigma^2_k$, and $\psi_k$.
Variance of observations uncertainty, $\sigma^2_\epsilon$, is assumed to
be known in this study.
This section motivates several choices of the sampling
strategy, while \ref{sec:samplingDetails} reportes the details of the sampling.

Sampling used the Block-at-a-Time Metropolis-Hastings
algorithm \citep{Hastings70, Chib95, Gelman03}. Each signal variance,
$\sigma^2_k$, was sampled from an inverse Gamma distribution. All the other
parameters were sampled by Metropolis steps. A Metropolis step can be used to
obtain a sample from a random variable for which probability density can be
evaluated up to a normalizing constant \citep{Metropolis53, Hastings70,
Andrieu03}. One Metropolis step was applied for each hyperparameter $\psi_k$
and another Metropolis step was applied for model parameter vector
$\boldsymbol{\btheta}$.
 Proposals of new parameter vectors were generated by using differential
evolution Markov chain algorithm \citep{terBraak08}.

We used a subset of observation locations, called the supporting
locations (triangles in Figure \ref{fig:GPEx_randomFunction}), for training the GP
model of discrepancies. The usage of a subset improved computational efficiency
and prevent numerical singularity, because the resulting matrices were smaller
and less prone to numerically singularity \citep{Brynjar14}.

In addition, we treated model discrepancies at supporting locations,
$\bdelta_s$, as latent variables instead of including them  in the set of
estimated parameters for two reasons. First, the spacing of the supporting
points should adapt to the correlation length to avoid numerical instabilities,
and hence the number and location of these supporting locations varied. Second,
this treatment dimished the problem of high rejection rate in the Metropolis
step and the slow mixing of the parameters, which occurs with sampling that is
conditioned on discrepancies \citep{Brynjar14}. Furthermore, the resulting
formulation can also be used with non-sampling based optimization methods, e.g.
gradient-based methods that require a smaller number of model evaluations.

\subsection{The basic example \label{sec:methodsExampleModel}}

A basic synthethic example demonstrates the problem with imbalanced multiple
data streams.

Consider the chemical reaction that oxidizes organic carbon in soil and evolves
carbon dioxide ($CO_2$). The $CO_2$ production rate was
monitored over a week, i.e.~roughly 1000 times, and found to increase with
temperature. For the same site the content of organic matter has been
measured september each year in 10 years together with the cumulative
amount of $CO_2$ produced over two weeks. The cumulated $CO_2$ production 
increased with organic matter content, i.e.~the amount of available reactant.

We modeled the corresponding two data streams $\mathbf{o}_{\mathrm{rich}}$
($n_{\mathrm{rich}}=1000$) and $\mathbf{o}_{\mathrm{sparse}}$
($n_{\mathrm{sparse}}=10$) of such a system by a surrogate physical process
\citep{Reich15}. The covariates of organic matter content
($\mathbf{x}_{\mathrm{sparse}}$) and temperature ($\mathbf{x}_{\mathrm{rich}}$)
were sampled from uniform distributions as $\sim \mathcal{U}(0.5, 1.5)$ and
$\sim \mathcal{U}(0.7, 1)$, respectively.

For demonstration purposes we use a simple linear model, that depends on
parameters $a$,$b$, and $c$ as, 
\numparts
\begin{eqnarray}
\label{eq:artExSparse}
\hat{\mathbf{o}}_{\mathrm{sparse}}(a,b) &= a \,
\mathbf{x}_{\mathrm{sparse}} + b \, \overline{x}_{\mathrm{rich}}/10.
\\
\label{eq:artExRich}
\hat{\mathbf{o}}_{\mathrm{rich}}(a,b,c) &= a \, x_{1,\mathrm{sparse}} + b
\, \left( \mathbf{x}_{\mathrm{rich}} - \mathrm{c} \right) 
.
\end{eqnarray}
\endnumparts

The example is set up in a way so that the the spread in the prediction of the
sparse data stream, $\mathbf{o}_{\mathrm{sparse}}$, mainly depends on
the covariates $\mathbf{x}_{\mathrm{sparse}}$. This is because the process
description (\ref{eq:artExSparse}) uses a single aggregated value
$\overline{x}_{\mathrm{rich}}$, of the covariates of the rich data stream,
specifically average september temperature.
The spread in the rich data stream $\mathbf{o}_{\mathrm{rich}}$, on the other
hand, mainly depends on covariates, $\mathbf{x}_{\mathrm{rich}}$.
This is because the process description (\ref{eq:artExRich}) only uses a single
value, $x_{\mathrm{1,sparse}}$, of the covariates of the rich data stream,
specifically the organic matter content at the year of the measurement
campaign.

The surrogate physical process, $\mathbf{o}^*$, was generated by running the
model with parameters $a^*=1$ and $b^*=2$, and bias variable $c^*=0.3$.
Next, observations, $\mathbf{o}$, were generated by adding Gaussian noise to
$\mathbf{o}^*$ with standard deviation of 4\% and 3\% of the mean of
$\mathbf{o}^*_{\mathrm{sparse}}$ and $\mathbf{o}^*_{\mathrm{rich}}$,
respectively. 

Using the generated observations and covariates, the posterior density of
parameters $a$ and $b$ were sampled by different scenarios of model inversions.
In order to demonstrate the transfer of model uncertainty, all scenarios used a
model that slightly differed from the data-generating model.
Specifically they used a value of the fixed bias parameter in the process
description of the rich data stream (\ref{eq:artExRich}) of $c=0.1$, instead of
$c^*=0.3$. This bias parameter represents a difference between measured air
temperature and the top soil temperature at the site of respiration.

\subsection{The real world ecosystem example
\label{sec:methodsDalecHowlad}}
The real world example inversely estimated 15 parameters and initial conditions
of the process-based ecosystem model of carbon dynamics, the Data Assimilation
Linked Ecosystem Carbon (DALEC) \citep{Williams05}.
Observations comprised 10-years of daily eddy covariance-based net ecosystem
exchange (NEE) \citep{Hollinger04, Hollinger05}, soil respiration, and litterfall at the
Howland forest.
While data streams of NEE and respiration had about 2000 observations,
litterfall had only one record for each of the 10 years. For demonstration
purposes, here, we used estimates of observation uncertainties that were
reduced by 25\% compared to the original estimates for each data stream.
Original estimates of observation error were specified as a function that
increased with the magnitude of the fluxes. We applied those functions to the
model prediction instead of the observed value in order to prevent preferential
fit to low fluxes. Further details of the model inversion settings are
described in \citep{Wutzler14}.

\section{Results \label{sec:results}}

First, the results of the basic example illustrate the consequences applying
several analyses that differ by their treatment of model discrepancy. Second,
the ecosystem example demonstrates the applicability of the GP approach to more
demanding real world inversion settings.

\subsection{Basic example  \label{sec:resultsBasicExample}}
\begin{figure}
\begin{center}\includegraphics{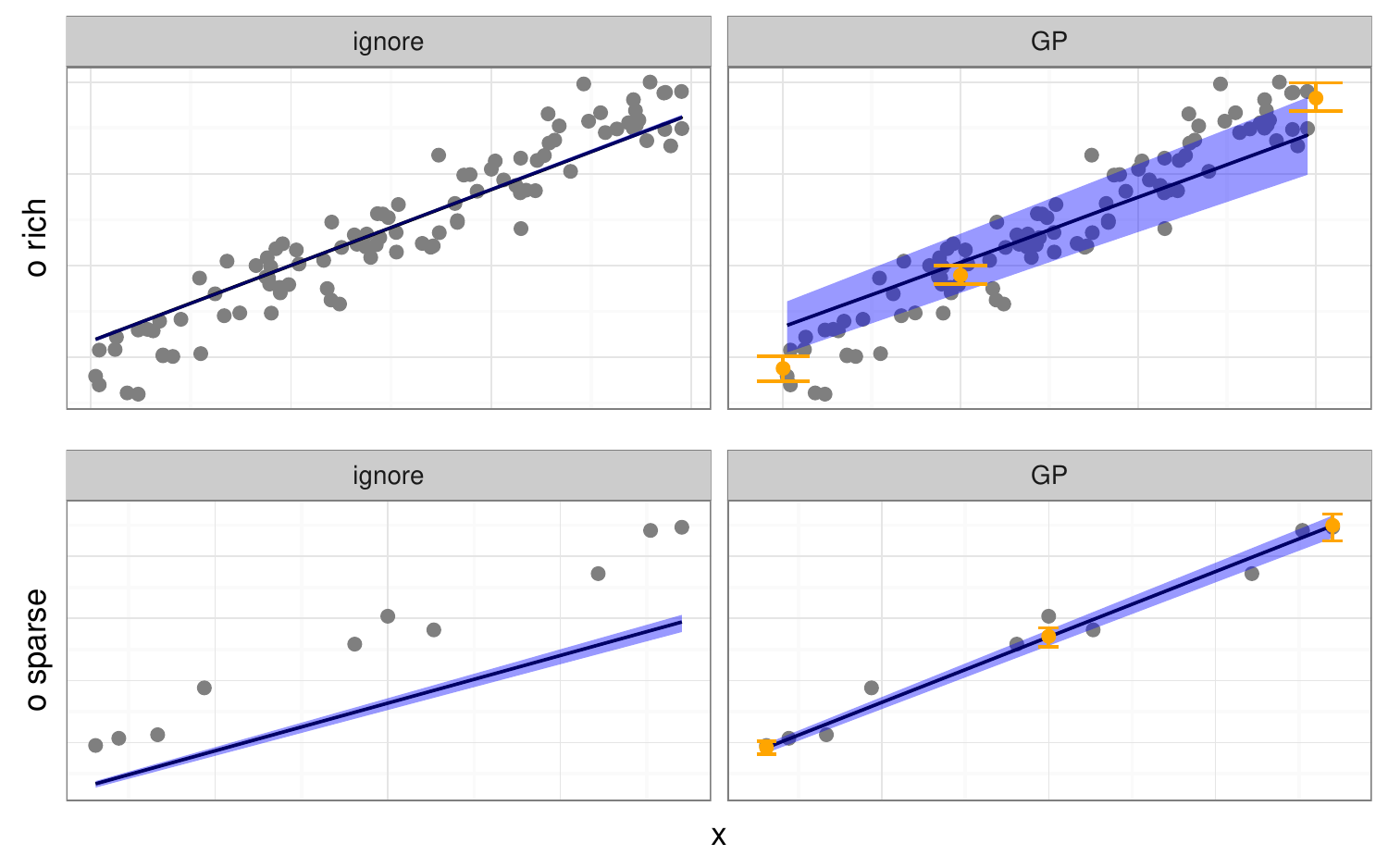}\end{center}
\caption{\label{fig:predPost} 
Predictive posterior of the basic example: Ignoring model discrepancy
(left) leads to allocation of model discrepancy to the sparse observations, as
seen by the misfit between observations (dots) and median model predictions
(line inside shaded 95\% confidence band).
 Contrary, representing discrepancies by a GP (right) correctly
allocates discrepancy to a slightly worse fit of the biased rich  observations.
The distribution of process predictions (vertical 95\% confidence whiskers)
hint to the form of the deficiency in the process formulation.
}
\end{figure}

The inversion of the basic example model (Section \ref{sec:methodsExampleModel})
 resulted in different estimates of posterior density of model parameters and in
 different estimates of predictive posterior density of process values (Figure
\ref{fig:predPost}) when using different sampling scenarios (Table
\ref{tab:invScen})

\begin{table}
\caption{\label{tab:invScen} Scenarios of model inversions.}
\lineup
\begin{tabular}{@{}*{3}{l}}
\br                              
Sampling scenario & Description  & Section \cr
\mr
ignore & ignoring model discrepancy & \ref{sec:ResIgnore} 
\cr
GP & GP for each stream's model discrepancy  &
\ref{sec:ResIndep}\cr
\br
\end{tabular}
\end{table}

\subsubsection{Analysis without accounting for model discrepancy
\label{sec:ResIgnore}}

With the ignore scenario, model discrepancy was assumed to be the zero vector,
$\bdelta = \mathbf{0}$. The Likelihood of the observations conditioned
on model parameters $\btheta$ was to the usual formulation
derived from the assumption of normally distributed observation
errors \citep{Tarantola05}.

\begin{eqnarray}
\label{eq:logThetaIgnore}
p_{\mathrm{indep}}(\btheta|\mathbf{o})
&= C_1 \, p(\mathbf{o}| \btheta)  \, p(\btheta).
\nonumber
\\
\mathrm{log}\left(p_{\mathrm{indep}}(\btheta|\mathbf{o})\right)
&= C_2 -1/2 \sum_k S_k(\mathbf{o}_k|\btheta) + \mathrm{log}\, p(\btheta).
\end{eqnarray}

\begin{eqnarray}
\label{eq:misfit2}
S_k(\mathbf{o}_k|\btheta, \bdelta) 
= \mathbf{d}^T \bSigma^{-1} \mathbf{d} 
= \sum_{i_k} \frac{ 
\mathbf{d}_{i_k}^2}{\sigma_{\epsilon,i_k}^2} ,
\\
\mathbf{d}_{k} = \mathbf{o}_{k} - \left[ \mathbf{g}_k(\btheta) +
\bdelta_{k} \right]
\end{eqnarray}

where $C_1$ and $C_2$ are constants, which cancel in the Metropolis decision,
and $i_k$ iterates across all record numbers in stream $k \in \{ rich, sparse
\}$, $S_k$ is a cost, and $\mathbf{d}_{k}$ is the vector of observation-process
residuals.

The infererred uncertainty of the process prediction was very low, indicated by
the thin 95\% confidence band of the model predictions (first column of Figure
\ref{fig:predPost}). Moreover, the introduced model deficiency
appeared in the sparse data stream instead of the rich data stream. 

By ignoring the model discrepancies, also the additional correlation among model
data residuals were ignored. Hence with this scenario, the uncertainty was largely
underestimated. Moreover, the results incorrectly suggested that the model
fails to predict the sparse data stream, and that the corresponding
process description should be refined. Instead, we had introduced the model
deficiency in the process description corresponding  to the rich data stream
(Section \ref{sec:methodsExampleModel}).

\subsubsection{Analysis with representing model discrepancy as a GP
\label{sec:ResIndep}}

With the GP scenario, model discrepancies were represented by a separate GP for
each stream.
Note that there was an additional term, $-1/2 \, \hat{\bdelta}_{s,k}
\mathbf{K}_{ss,k}^{-1} \hat{\bdelta}_{s,k}$,
 in the log-density of the parameters (\ref{eq:LogLikTheta} -
 \ref{eq:LogDenDelta}) compared to the ignore scenario
 (\ref{eq:logThetaIgnore}), effectively penalizing large model discrepancies.

The infererred uncertainty of the process corresponding to the rich
observations increased compared to the ignore scenario (Figure
\ref{fig:predPost}).
The magnitude of model discrepancies in the sparse data stream strongly
declined, while the magnitude of model discrepancies in the rich data stream
only slightly increased.

The GP scenario helped to tackle both problems: first, the underestimation of
posterior variance and second, the unbalanced allocation of model
discrepancies.

\subsubsection{Analysis using a gradient-based optimization
\label{sec:ResGradient}}
The optimum parameter set can be found by a gradient-based maximization of
the probability densitiy (\ref{eq:logThetaIgnore}). Here, we used the
Broyden-Fletcher-Goldfarb-Shanno (BFGS) method of \citep{Nash90}, as implemented
by the \textit{optim} function in the R \textit{stats} package \citep{R07}. A
first order estimate of parameter uncertainty was obtained from the Hessian
matrix at the optimum.
 
\begin{figure}
\begin{center}\includegraphics{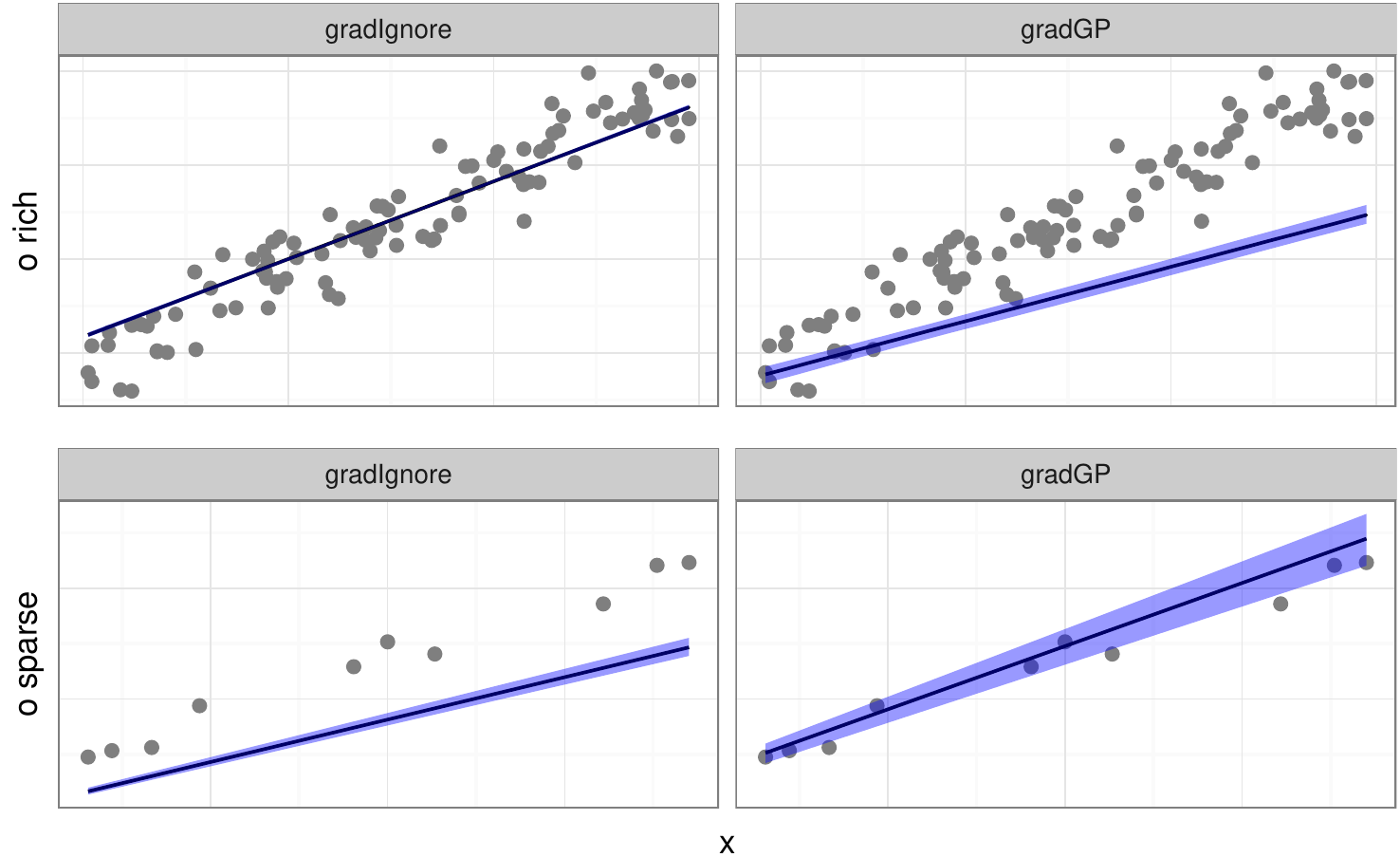}\end{center}
\caption{\label{fig:predPostGrad} 
Predictive posterior based on a gradient-based
optimization of the basic example reveals similar changes on applying the
GP-based formulation as with the sampling based inversion (Figure
\ref{fig:predPost}).
The GP-based inversion (right) correctly
allocates discrepancy to the biased rich observations, 
as seen by the misfit between observations (dots) and median model
predictions (line inside shaded 95\% confidence band).
}
\end{figure}

Alternatively, the optimum parameter set can be found by maximization of the
GP-based density (\ref{eq:LogLikTheta}). In order to maximise this
density, we had to a priori specify the parameters of the GP, i.e.
supporting locations, correlation length, $\psi_k$, and signal
variance, $\sigma^2_{d,k}$. We specified conservative values based on the
results of the optimization that ignored discrepancies (Figure \ref{fig:predPostGrad}
left), where the magnitudes of the model discrepancies exceeded the observation
uncertainty, while the correlations spanned the entire range of the data.
Specifically, we specified four equidistant supporting points, a correlation
length of one-third of the range of the respective covariate, and a signal
variance based on 1.5 times the observation uncertainty for all data streams. 

The results with gradient-based optimization changed in a similar way as in the
sampling-based inversions when representing model discrepancy by a GP. The
estimates of the prediction uncertainty increased, while the model discrepancy
in the sparse data stream almost disappeared (Figure \ref{fig:predPostGrad}).

\subsection{Real world ecosystem example  
\label{sec:resultsDalec}}

Model inversions of the 15 DALEC parameters using datastreams with
thousands of records was computationally more expensive, both in terms of
required model runs and in terms of computing the inverse of the discrepancy
correlation matrix. Despite the improved GP sampling formulation, mixing was not
as good as expected and we had to increase the thinning interval up to 48. The slowest
mixing was observed for correlation length parameters. Chains converged to the
limiting distribution after about 2000 samples (\ref{sec:DalecMoreResults}).

\begin{figure}
\begin{center}\includegraphics{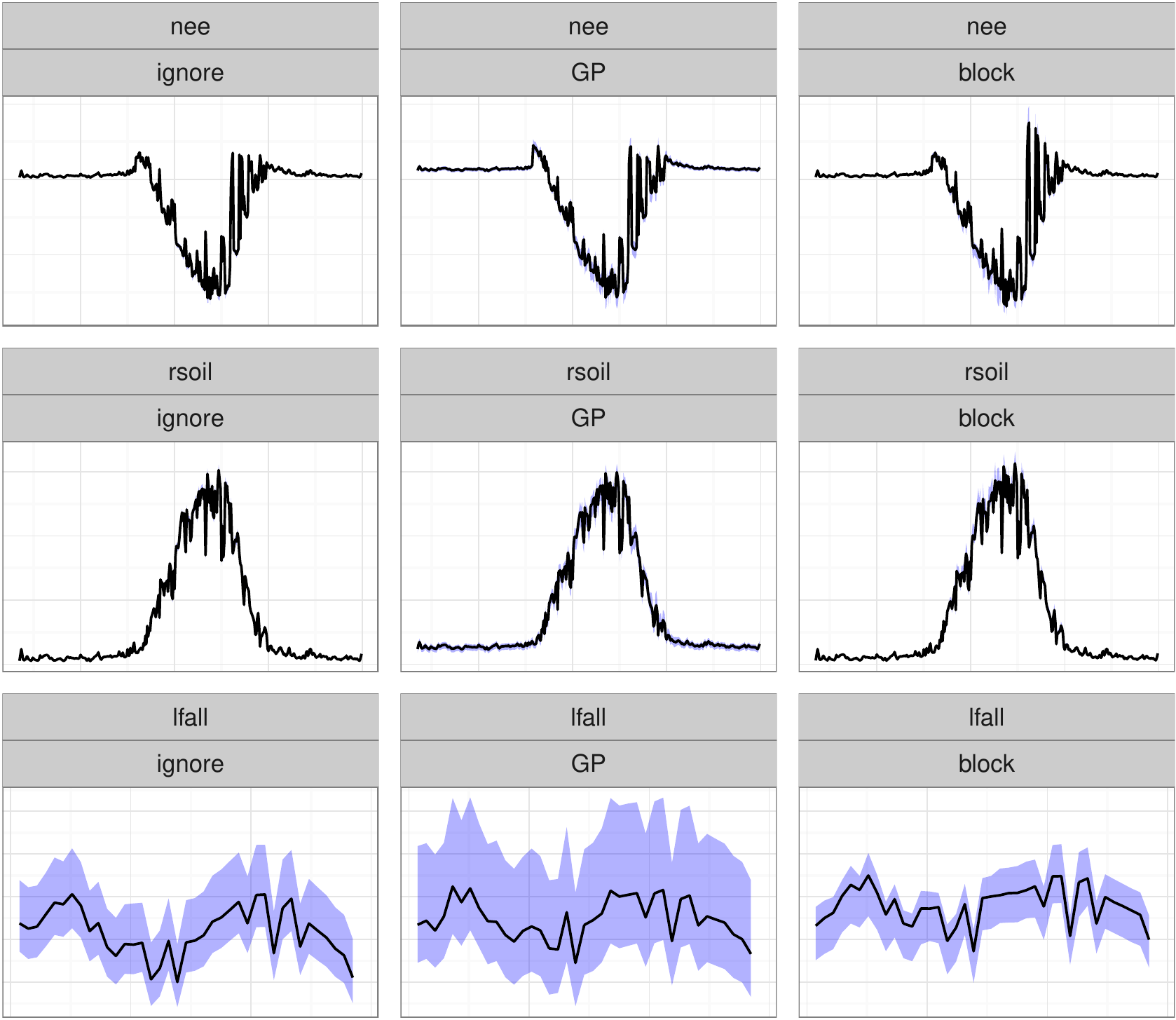}\end{center}
\caption{\label{fig:HowlandPredPostFuture} 
Long term (40 years) future predictive posterior of the DALEC
ecosystem example show a more realistic
estimation of prediction uncertainty (shaded
95\% confidence bands) with the GP and the block approach.
Ignoring discrepancy (left) leads to overconfident estimates, as seen by the
overplotting of the confidence interval by the median model prediction in the
rich NEE and respiration data streams (two top rows, where only the 40\textsuperscript{th}
year is shown).
While the GP approach (middle column) leads to increased uncertainty of both rich and sparse
data streams, the block approach (right) balances the allocation of
uncertainty away from the sparse litterfall stream (bottom row) towards the rich
data streams.
} 
\end{figure}

When ignoring model discrepancies, posterior distribution of parameters were
estimated in such a way that predictions beyond the calibration period lead to
very low uncertainty of predictions of the rich data streams of NEE and
soil respiration (first column in Figure
\ref{fig:HowlandPredPostFuture}).
With the GP scenario the predicted uncertainty increased (second column in
Figure \ref{fig:HowlandPredPostFuture}).
This was similar to the results of an inversion based on the parameter blocks
approach \citep{Wutzler14} (shown for comparison in third column of Figure
\ref{fig:HowlandPredPostFuture}).

The marginal parameter distributions estimated by the GP scenario were mostly
broader than the estimates of the ignore and the blocks
scenario. Moreover, the GP scenario estimated higher turnover rates (Tl, Tf,
Tlab) and lower temperature sensitivity (Et) (\ref{sec:DalecMoreResults}). 

\begin{figure}
\begin{center}\includegraphics{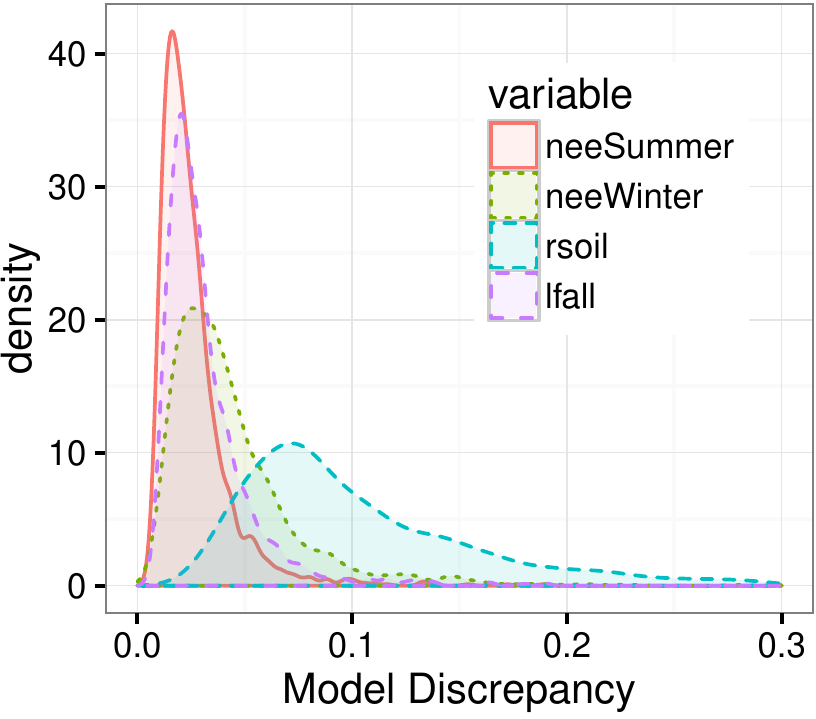}\end{center}
\caption{\label{fig:HowlandModelDiscrepancy}  
Distribution of discrepancy variance in Howland 
example was only a small fraction of corresponding variance of observation
uncertainty (see Equation \ref{eq:sigmaK}). Such a finding increases confidence
in the application of the DALEC model.
}
\end{figure}

The GP inversion estimated the largest model discrepancies for soil
respiration (Figure \ref{fig:HowlandModelDiscrepancy}), at about double the
discrepancy of all the other data streams. The overall magnitude was, however, small with only less
than 10\% of average observation uncertainty of the corresponding
data stream. The DALEC model well captured the patterns in the observations and
there was only a negligible trade-off between several constraining data streams.

\section{Discussion \label{sec:discussion}}

The discussion highlights three main points. Section
\ref{sec:discussionMultiple} discusses how the GP approach can solve problems associated with imbalanced data
streams.
Next, section \ref{sec:discussionTechnical} discusses how methodological
innovations allowed the application to real world problems.
Finally, section \ref{sec:discussionOptimizer} highlights, that the proposed
formulation allows combining advantages of non-sampling based optimizations
with GP-based representation of model discrepancy.

\subsection{Balancing multiple data streams
\label{sec:discussionMultiple}}

Model discrepancy will become more relevant in future. It is often caused by a
mismatch in detail between model and observations. The detail in models is
bounded by the modeling purpose, while the detail in observations increases
with the ability to take high resolution measurements \citep{Luo11a}.

Therefore, the problem of preferential allocation of model discrepancy to sparse
data streams \citep{Wutzler14} should be acknowledged when inverting models
against imbalanced multiple data streams. It impairs the usage of multiple data
streams to constrains different model aspects \citep{Richardson10}, as
demonstrated, here, with the basic example. Most important, it complicates
locating the source of model deficiencies, because the model discrepancy does
not appear with the observations corresponding to the weak model aspects (Figure
\ref{fig:GPEx_posterior}).

The GP approach, i.e.~explicit representation of model discrepancy by a GP
(Figure \ref{fig:GPEx_randomFunction}), tackles the problem of preferential
discrepancy allocation. The problem is caused in part by overestimating the
information content in the rich data stream, which is in turn caused by not
accounting for correlations among model-data residuals due to model discrepancy.
The GP approach represents these correlations by two parameters per data stream,
whose distribution is estimated during the Bayesian model inversion. Indeed, in
this study it achieved a better balance of fits to the imbalanced data streams
(Figure \ref{fig:predPost}). In addition, it yielded a more realistic estimate
of the uncertainty of the parameters and the predictions.

\begin{figure}
\begin{center}\includegraphics{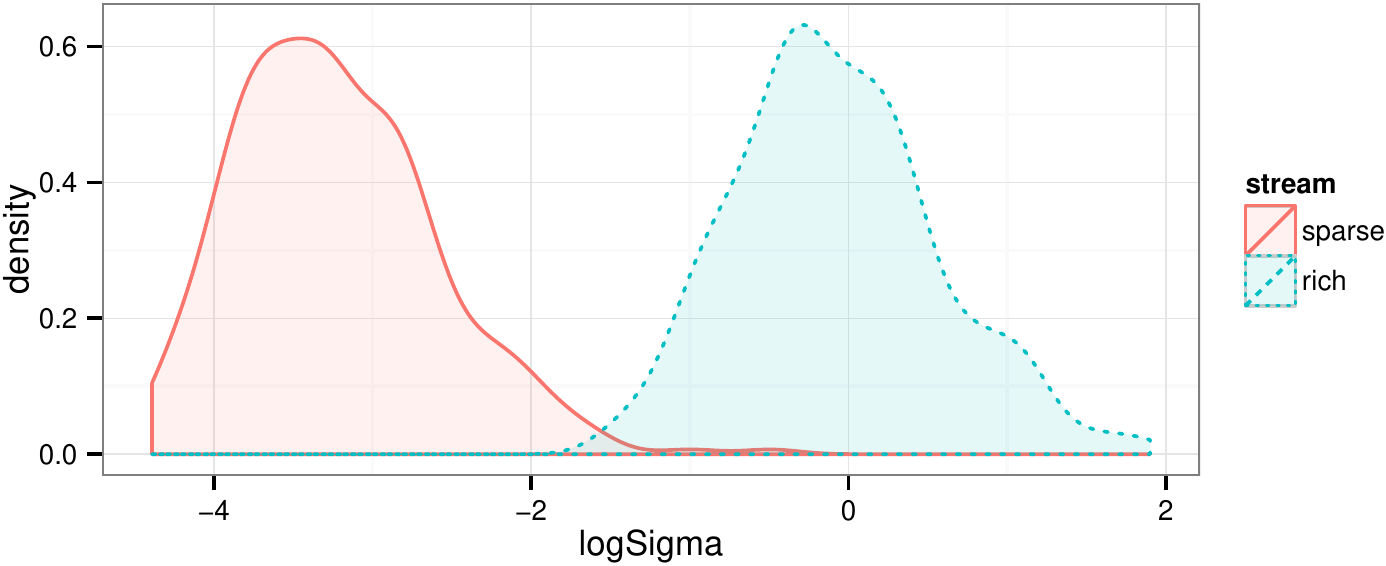}\end{center}
\caption{\label{fig:signalVariance}
The large values in the distribution of posterior model discrepancy, 
which are of the same magnitude as the variance of observation
uncertainty (units of the x-axis is as natural logaritm of their ratio, see
equation \ref{eq:sigmaK}), suggest that model developers should 
refine the process description predicting the rich 
data stream with the basic example.
}
\end{figure}

The estimated variance of the model discrepancies can be used to identify which
processes in the model need refinement. The variance can be expressed in a
dimensionless metric, normalized as a multiple of average stream measurement
uncertainty (section \ref{sec:dimensionlessSigma}), which is not affected by
different units.
A larger normalized variance (relative to other streams) indicates that the
corresponding process are in less accordance with observations. In the basic
example, the largest normalized misfit was correctly associated with the process
that predicts the observations of the rich data stream (Figure
\ref{fig:signalVariance}), as the example was constructed.

Model refinement can be justified by the magnitude of the normalized
discrepancy variance. Refinement is necessary up to a detail that captures the salient
patterns in the observations \citep{Wiegand03}. Therefore, it is justified when
the data is good enough to reveal more details of the processes, and when those
details help the modeling purpose. With the Howland case, refinement was
not necessary, because, the model discrepancy was only a small fraction of
the observation uncertainty (Figure \ref{fig:HowlandModelDiscrepancy}).
With the basic example, refinement was justified, despite the cost of an
additional parameter, because model discrepancy variance of the rich data stream
amounted to same magnitude as the observation uncertainty (Figure
\ref{fig:signalVariance}, $e^0=1$). In addition, refinement can be guided by 
the shape of the model discrepancy. With the basic example, refinement would
start to look at the slope in the modeling of the rich data stream (Figure
\ref{fig:predPost}).


The GP approach propagates uncertainty in observations to the uncertainty of
parameters in a well-grounded manner. It is fully based on probabilistic
principles, and does not invoke any external information for combining multiple
criteria.
It avoid the modification of uncertainty estimates with weighting different data
stream \citep{Wutzler14}. It retains the advantage that
improved resolution of sampling also leads to a more precise estimation of model
parameters, given that the model is able to resolve the patterns emerging at the
high resolution.

The acceptance of more parameter vectors with the GP-approach compared to the
approach ignoring discrepancy is feature not a flaw. The resulting higher
estimates of parameter and prediction uncertainty are more realistic, because
there is less information in model-data residuals that are correlated due to
model discrepancy. But how can the approach decide in which shares the
model-data residuals are explained by either the mechanistical model or the
statistical GP model of discrepancies? The allocation of model-data misfit, to
either the mechanistical model or the statistical GP model, is solved by
penalizing large model discrepancies (\ref{eq:LogDenDelta}). Parameter vectors
that yield predictions with low discrepancy lead to low estimates of the
signal variance and in turn to high penalties and low probabilities for other
parameter vectors.

\subsection{Application to larger data streams
\label{sec:discussionTechnical}}

While the GP approach worked well for the basic example, there were practical
problems with application to larger problems, which were tackled by the proposed
sampling scheme.

The first problem is that an inversion using rich data streams with several
thousand records \citep{Luo11a} must use many supporting points for the GP if
correlation length becomes small. That leads to the construction of huge
matrices that need to be inverted often. We tackled this problem by constraining
the choices of supporting points and correlation length (\ref{sec:GPLimits}).

The second problem is the slow mixing of model parameters and model
discrepancies when parameters and discrepancies are sampled separately
\citep{Brynjar14}.
In this study, we tackled this problem by conditioning the model discrepancies
on observed model-data residuals, i.e.~treating them as latend variables,
and recomputing them on each draw of new model parameters or correlation length.
The posterior density of model parameters then is a part of the joint
likelihood of model parameters and inferred discrepancies. This derivation
resulted in an additional term per data stream in the density of model
parameters that penalizes large model discrepancies (\ref{sec:samplingDetails}).

The third problem is to specify appropriately spaced supporting points
(triangles in figure \ref{fig:GPEx_randomFunction}) if correlation length $\psi$
is not well known a priori. If the points are spaced to close to each other, the
matrices become numerically singular \citep{Brynjar14}. If the points are spaced
too wide, the underlying function may not be captured well. We tackled this
problem by re-arranging the supporting points for each proposed correlation
length. In addition, we introduced some randomness in selecting supporting
points for a given correlation length. This avoided being trapped in states with
lucky choices of supporting locations that yield exceptionally high likelihoods
of observations.

The proposed sampling scheme tackled those problems and allowed
successfull application to real world data of several thousand records. 

\subsection{Combination with fast optimizers
\label{sec:discussionOptimizer}}

The proposed formulation (\ref{eq:LogLikTheta} - \ref{eq:LogDenDelta}) makes it
possible to combine the advantages of gradient-based optimization with GP-based
representation of model discrepancy.

Speed is one advantage of gradient-based optimization compared to Bayesian
sampling.  The proposed Bayesian sampling scheme converges faster and is
numerically less demanding than a sampling scheme that needs to sample
discrepancies at the supporting locations. However, many real world problems
employ models with longer run-time and cannot affort Monte-Carlo approaches, but
must rely on gradient-based optimization for model calibration.

With gradient-based optimization, the parameters of the GP could not be
estimated from the data any more. They had to be specified a priory based on reasonable
expectations about the magnitude of the model discrepancy, i.e.~signal variance,
and the smoothness of the discrepancy, i.e.~the correlation length.
The advantages of the GP-based representation held true, despite the
specification of a conservatively large signal variance for the sparse data
stream with the basic example (Section \ref{sec:ResGradient}).

A first order estimate of uncertainty in model parameters can be obtained from
the curvature, i.e.~Hessian matrix at the optimum of a gradient-based solution.
Whenever computationally possible this estimate should be improved in a second
step by a Bayesian sampling scheme. In this second step also the GP parameters
can be estimated from the data as their uncertainty contributes to the
uncertainty of parameters and predictions.


\section{Conclusions}
Based on results of different sampling scenarios of the basic model example
and of the ecosystem case we conclude:

\begin{itemize}
  \item Neglecting model discrepancies during the sampling leads to, first, 
   underestimation of posterior uncertainty, and second, preferential 
  allocation of model discrepancy to sparse data streams.
  \item Explicitly accounting for model discrepancy by representing it
  as a Gaussian processes (GP) successfully tackles both problems.
  \item The proposed sampling scheme, which treats model discrepancies as
  latent variable, tackles several computational
  problems occuring with large data streams. It allows application of the
  GP approach to real world inverse problems.
  \item The proposed formulation can combine advantages of
  gradient-based optimization with GP-based representation of model
  discrepancy.
\end{itemize}

When inverting a model using multiple data streams, it is important to
explicitly account for model discrepancies. The presented GP formulation
efficiently balances allocation of model discrepancy to imbalanced data streams
and allows an improved inference on modelled processes.

\appendix
\section{Sampling details \label{sec:samplingDetails}}

\begin{table}
\caption{Notation. Stream subscript $k$ is omitted
for brevity. \label{tab:notation}}
\begin{tabular}{@{}*{2}{l}}
\br                              
Symbol & description \cr
\mr
$\mathbf{o}$ & observations. vector of size $n_k$ \cr

$\btheta$ & model parameters \cr
$\mathbf{g}(\btheta)$ & vector of model prediction for each observation\cr 

$\mathbf{t}$ & locations of the observations, here their times. vector of
size $n_k$ \cr 

$\mathbf{z}$ & model-data residuals. $\mathbf{z} = \mathbf{o} -
\mathbf{g}(\btheta) = \mathbf{z}_s \cup \mathbf{z}_r$\cr

$\mathbf{z}_s$ & model-data residuals at supporting locations $\mathbf{s}
\subset \mathbf{t}$.
\cr

$\mathbf{z}_r$ & model-data residuals at remaining locations $\mathbf{r} =
\mathbf{t} \setminus \mathbf{s}$. \cr

$\mathbf{d}$ & model-process residuals. $\mathbf{d} = \mathbf{o} -
\left[ \mathbf{g}(\btheta) + \bdelta \right]$\cr

$\psi$ & correlation length of model discrepancies \cr 

$\sigma^2_\epsilon$ & variance of observation uncertainty. \cr

$\sigma^2_d$ & variance of model discrepancy, i.e.~signal variance of
the Gaussian process (GP).
\cr

$\sigma^2$ & normalized variance of model discrepancies.
$\sigma^2 = \sigma^2_d / \sigma^2_\epsilon$ (\ref{eq:sigmaK}) \cr

$\bdelta$ & model discrepancy. $\bdelta= \bdelta_s \cup \bdelta_r$ \cr

%

$\bLambda_{\mathbf{i}, \mathbf{j}}$ & Correlation matrix between two vectors of
locations $\mathbf{i}, \mathbf{j} \subseteq \mathbf{t}$. It depends on $\psi$ \cr 

$\mathbf{K}_{\mathbf{i}, \mathbf{j}}$ & Covariance matrix between two vectors of
model locations. 
$\mathbf{K}_{\mathbf{i}, \mathbf{j}} = \sigma^2_d \bLambda_{\mathbf{i},
\mathbf{j}}$
\cr

$\mathbf{K}_z$ & Covariance matrix with observation noise.
$\mathbf{K}_z = \mathbf{K}_{\mathbf{s}, \mathbf{s}} +
\sigma^2_\epsilon \mathbf{I}$ \cr 

\br
\end{tabular}
\end{table}

The following section details the sampling of parameters for one data stream. 
 All quantities except model parameters, $\btheta$, are stream specific.
For brevity the stream index $k$ is omitted.

A complete formulation of the model is the following
\begin{eqnarray}
\mathbf{o} | \btheta, \bdelta  \sim 
N\left(g(\btheta) + \bdelta, \sigma^2_\epsilon \mathbf{I} \right)
\\
\left[\btheta\right] \propto 1 
\\
\bdelta | \psi, \sigma^2_d \sim GP \left( \mathbf{0}, K(\psi, \sigma^2_d)
\right)
\\
\psi \sim \mathrm{trGamma}(a_\psi,b_\psi) 
\\
\sigma^2_d = \sigma^2 \sigma^2_{\epsilon} 
\,, \sigma^2 \sim \mathrm{IG}(\alpha_{\sigma^2},\beta_{\sigma^2}) 
,
\end{eqnarray}

\noindent
where, in the presented examples the prior for model parameters $\btheta$ was
uniform, and $\mathrm{trGamma}$ denotes the truncated gamma distribution.
The covariance function of the GP, here, was a squared exponential function:
$K(x_p,x_q; \psi, \sigma^2_d) = \sigma^2_d exp(-(x_p-x_q)^2/\psi^2 +
\sigma^2_\epsilon \delta_{pq})$, where the Kronecker delta $\delta_{pg}$ equals
one for $p=q$ and zero otherwise.
In this study we knew that the correlation length of model discrepancies had to
be of similar magnitude as the range of observation locations and used a prior
to yield a mean of one third of the range of locations, $r(\mathbf{x}_k)$, and
variance of $r(\mathbf{x}_k)^2/3.2$, specifically $(a_\psi,b_\psi) =
(1.14,0.188)$ for the sparse data stream. In order to avoid numeric instability,
the Metropolis step of $\psi$ rejected proposals if they fell outside the
truncation bounds described in \ref{sec:GPLimits}. The prior density of the
normalized discrepancy variance is an Inverse Gamma distribution. We used
$\alpha_{\sigma^2}=1.005$ and $\beta_{\sigma^2}=0.1$ to yield a rather flat
prior distribution with mean 20 and mode 0.5.

The Block-at-a-Time Metropolis-Hastings algorithm proceeds by several cycles
with each cycle in turn sampling several blocks, i.e.~subsets of the parameter vector
\citep{Hastings70, Chib95, Andrieu03}.
Here we use a Metroplis sampling block for each $\psi_k$, another Metropolis
sampling block for $\btheta$, and Gibbs sampling blocks for each
$\sigma^2_{d,k}$.
We briefly recall the Metropolis-Markov chain procedure to sample a random
variable for which the probability density ist known up to a constant.
At the beginning of each Metropolis sampling block a new parameter proposal is
generated.
In this study we generated proposals by using DEMC, which suggests steps
in parameter space based on the distribution the parameter among several
sampling chains \citep{terBraak08}.
Next, model discrepancies, and the conditional probability for proposed
block parameters are computed. These are also re-computed for the current
block parameters if parameters that are used by the probability function have
been updated by other blocks. Next a Metropolis decision accepts the
proposed state if the ratio of the probabilties (proposed to current) is larger than a random number drawn
from $U(0,1)$. If accepted, the proposed parameter is recorded as new sample,
otherwise the current parameter is recorded again.

Correlation length $\psi$ is sampled by a Metropolis-Hastings block. The full
conditional distribution of $\psi$ is
\begin{eqnarray}
\label{eq:pPsiCond}
p \left( \psi | \mathbf{o}  \right)
= \frac{ p \left( \psi, \hat{\bdelta} | \mathbf{o}  \right) }{ p \left(
\hat{\bdelta} | \psi, \mathbf{o} \right) } 
\propto \frac{ p(\mathbf{o}| \psi, \hat{\bdelta} ) \,
p(\psi, \hat{\bdelta}) }{ p \left(
\hat{\bdelta} | \psi, \mathbf{o} \right) }
= \frac{ p(\mathbf{o}| \psi, \hat{\bdelta} ) \,
p(\hat{\bdelta}|\psi) p(\psi) }{ p \left(
\hat{\bdelta} | \psi, \mathbf{o} \right) }
\\
\label{eq:LogLikObs}
p(\mathbf{o} | \psi, \hat{\bdelta}) \propto exp \left( -\frac{1}{2
\sigma^2_\epsilon} \| \mathbf{o} - \left( \mathbf{g} + \hat{\bdelta} \right)
\|^2 \right)
\\
p(\psi) = p_\Gamma(a_\psi,b_\psi) \propto \psi^{a_\psi-1} \, e^{ -b_\psi \psi} 
\\
\label{eq:LogLikDelta}
\frac{ p(\hat{\bdelta}|\psi) }{  p \left(
\hat{\bdelta} | \psi, \mathbf{o} \right) }
= \frac{(2 \pi)^{-n_k/2} |\mathbf{K}_{\bdelta}|^{-1/2} e^{ -\frac{1}{2} \,
(\hat{\bdelta} -\mathbf{0})
\mathbf{K}_{\bdelta}^{-1} (\hat{\bdelta}-\mathbf{0})}  }{(2 \pi)^{-n_k/2}
|\mathbf{K}_{\bdelta}|^{-1/2} \, e^{0}} \approx e^{ -\frac{1}{2} \,
\hat{\bdelta}_s \mathbf{K}_{ss}^{-1} \hat{\bdelta}_s }
\\
\mathbf{K}_{\bdelta} = \left[ \mathbf{K}_{ss}, \mathbf{K}_{sr};
\mathbf{K}_{rs}, \mathbf{K}_{rr} \right]
,
\end{eqnarray} 

\noindent
where the depencies on $\btheta$ and $\sigma^2_d$ and the stream index, $k$, are
omitted for brevity. The first step in (\ref{eq:pPsiCond}) derives from a
factorization of the joint density of $\psi$ and $\hat{\bdelta}$. The second
step is the Bayes rule. The third step is the factorization of the joint
unconditional density of $\hat{\bdelta}$ and $\psi$.

The Likelihood \ref{eq:pPsiCond} is based only on the expected value of model
discrepancy and does not require a sample of model discrepancy. It actually
holds for any particular sample of model discrepancy $\bdelta$, but choosing the
expected value $\hat{\bdelta}$ greatly simplifies calculations
(\ref{eq:LogLikDelta}).
The normalizing factor in (\ref{eq:LogLikDelta}) cancels. The numerator is the
probability density of $\hat{\bdelta}$ without knowning the observations, i.e.~a
multivariate normal density of the GP with zero mean. The denominator is the
probability conditioned on the current observations and predictions, i.e.~a
multivariate normal density with mean $\hat{\bdelta}$. The inversion of the
blocked matrix $\mathbf{K}_{\bdelta}$ could be done using blockwise inversion.
However, it still requires an inversion of a matrix of dimension of the number
of non-supporting locations $r$, and has the danger of becoming numerically
singular. We suggest approximating the norm by only using the supporting
locations. Due to the relation between spacing of supporting locations and the
correlation length (\ref{sec:GPLimits}), the approximation stayed within one
per mill precision in our applications so far.
The relation also prevents $\mathbf{K}_{\mathbf{s},\mathbf{s}}$ from becoming
numerically singular. If $\mathbf{K}_{\mathbf{s},\mathbf{s}}$ became numerically
singular, a small diagonal component could be added.

The vector of expected value of model discrepancies $\hat{\bdelta} =
\hat{\bdelta}_s \cup \hat{\bdelta}_r$ is computed as function of parameters
$\btheta$, $\psi$, and $\sigma^2_d$ based on the model model-data residuals at
supporting
 locations, $\mathbf{z}_s$ \citep{Rasmussen06}
\begin{eqnarray}
\label{eq:deltaS} 
\label{eq:bDeltaS} 
\hat{\bdelta_s} = \mathbf{K}_{\mathbf{s},\mathbf{s}}
\mathbf{K}_z^{-1} \mathbf{z}_s
\\
\label{eq:deltaR}
\hat{\bdelta_r} = \mathbf{K}_{\mathbf{r}, \mathbf{s}}
\mathbf{K}_{\mathbf{s},\mathbf{s}}^{-1} \bdelta_s = \mathbf{K}_{\mathbf{r},
\mathbf{s}} \mathbf{K}_z^{-1} \mathbf{z}_s
\end{eqnarray}
Note that covariance matrices $\mathbf{K}$ depend on current
discrepancy variance $\sigma_d^2$ and correlation length $\psi$, and that $\mathbf{z}_s$
depends on parameters $\btheta$ and observations $\mathbf{o}$.

Model parameters $\btheta$ are sampled by Metropolis-Hastings block.
Their conditional distribution depends on all data streams, $k$. We
assume that observations and model discrepancies are independent between
different streams.
\begin{eqnarray}
\label{eq:LogLikTheta}
p \left( \btheta | \mathbf{o}  \right)
& =  \frac{p \left( \btheta, \hat{\bdelta} | \mathbf{o} \right)}
   { p \left( \hat{\bdelta} | \btheta, \mathbf{o} \right) }
\propto \frac{  p(\mathbf{o}| \btheta, \hat{\bdelta}) \,
 p(\hat{\bdelta}|\btheta) \, p(\btheta) }
 { p \left( \hat{\bdelta} | \btheta, \mathbf{o} \right) }
 \\
&= \prod_k   p(\mathbf{o}_k| \btheta, \hat{\bdelta}_k) \, \frac{
 p(\hat{\bdelta}_k|\btheta) }
 { p \left( \hat{\bdelta}_k | \btheta, \mathbf{o}_k \right) }
 \, p(\btheta)
\\
p(\btheta) &\propto 1
\\
\label{eq:LogDenDelta}
\frac{ p(\hat{\bdelta}_k|\btheta) }{ p \left( \hat{\bdelta}_k | \btheta, \mathbf{o}_k \right) }
&= exp( -1/2 \, \hat{\bdelta}_k
\mathbf{K}_{\bdelta,k}^{-1} \hat{\bdelta}_k ) 
\approx
exp( -1/2 \, \hat{\bdelta}_{s,k} \mathbf{K}_{ss,k}^{-1} \hat{\bdelta}_{s,k} )
, 
\end{eqnarray}

\noindent
where, the dependence on all hyperparameters $\psi_k$ and $\sigma^2_{d,k}$ has
been omitted for brevity. The derivation of each stream-factor is analogous to
the derivation of conditional density $p \left( \psi | \mathbf{o}  \right)$
above, unless the simplification that the normalizing factor of
$p(\hat{\bdelta}_k)$ does not depend on $\btheta$. Note that a penalty term
(\ref{eq:LogDenDelta}) for each stream model discrepancy must be included in the
conditional density function of the parameter vector.
Again, we approximated the norm of the vector of all model discrepancies
$\hat{\bdelta}_k$ by the norm of the model discrepancies at supporting locations
$\hat{\bdelta}_{s,k}$.

Normalized discrepancy variance $\sigma^2$ was sampled from an inverse Gamma
distribution conditioned on discrepancies derived for current parameters.
\begin{eqnarray}
\label{eq:pSigma2Cond}
\sigma^2 | \mathbf{o}, \btheta, \psi
\sim IG \left(  \alpha_{\sigma^2} + \frac{n_s}{2},  \beta_{\sigma^2} +
\frac{1}{2 \sigma^2_{\epsilon}} ||
\hat{\bdelta}_s(\mathbf{o}, \btheta, \psi, \sigma^2_d) ||_{\Lambda_{ss}^{-1}}
\right) ,
\end{eqnarray}
where $n_s$ is the number of supporting locations and $\alpha_{\sigma^2}$,
$\beta_{\sigma^2}$ are parameters of the prior probability density. 
The factor $1/\sigma^2_{\epsilon}$ needs to be included, because we specified 
the prior for the discrepancy variance normalized by variance
 of observation uncertainties.
We did not encounter problems when reusing the current value of $\sigma^2_d$ in
the calculation of discrepancy. However, one could instead use a data-based
discrepancy variance to prevent this dependence and potential feedback
behaviour.
A data-based estimate can be found by maximising 
the likelihood of observed residuals given all the other
parameters and observation uncertainty.

\section{Limits for choosing supporting points $\mathbf{s}$ and
truncation  of sampling correlation length $\psi$ \label{sec:GPLimits}}

For large correlation length $\psi$, some matrices used in the calculation 
of model discrepancies $\delta_s$ become numerically singular.
Further, different correlation length that are larger than the range
of the data, $t$, cannot be distinguished. Hence, an upper bound of $(max(t) -
min(t))$ is applied to the sampling of $\psi$.

For very small correlation length, model discrepancy goes to the expected value
zero between supporting points. However, we want to model a smooth discrepancy
between supporting points. Hence, a lower bound of 2/3 of the mean distance
between supporting points is applied to the sampling of $\psi$. 

Supporting points are chosen among observation points closest to a grid with
distance $3 \psi /2 $, with a minimal spacing so that there are two points
between supporting points and a maximal spacing so that there are at least five
supporting points.

\section{Dalec-Howland inversion additional results 
\label{sec:DalecMoreResults}} 

\begin{figure}
\begin{center}\includegraphics[totalheight=0.95\textheight]{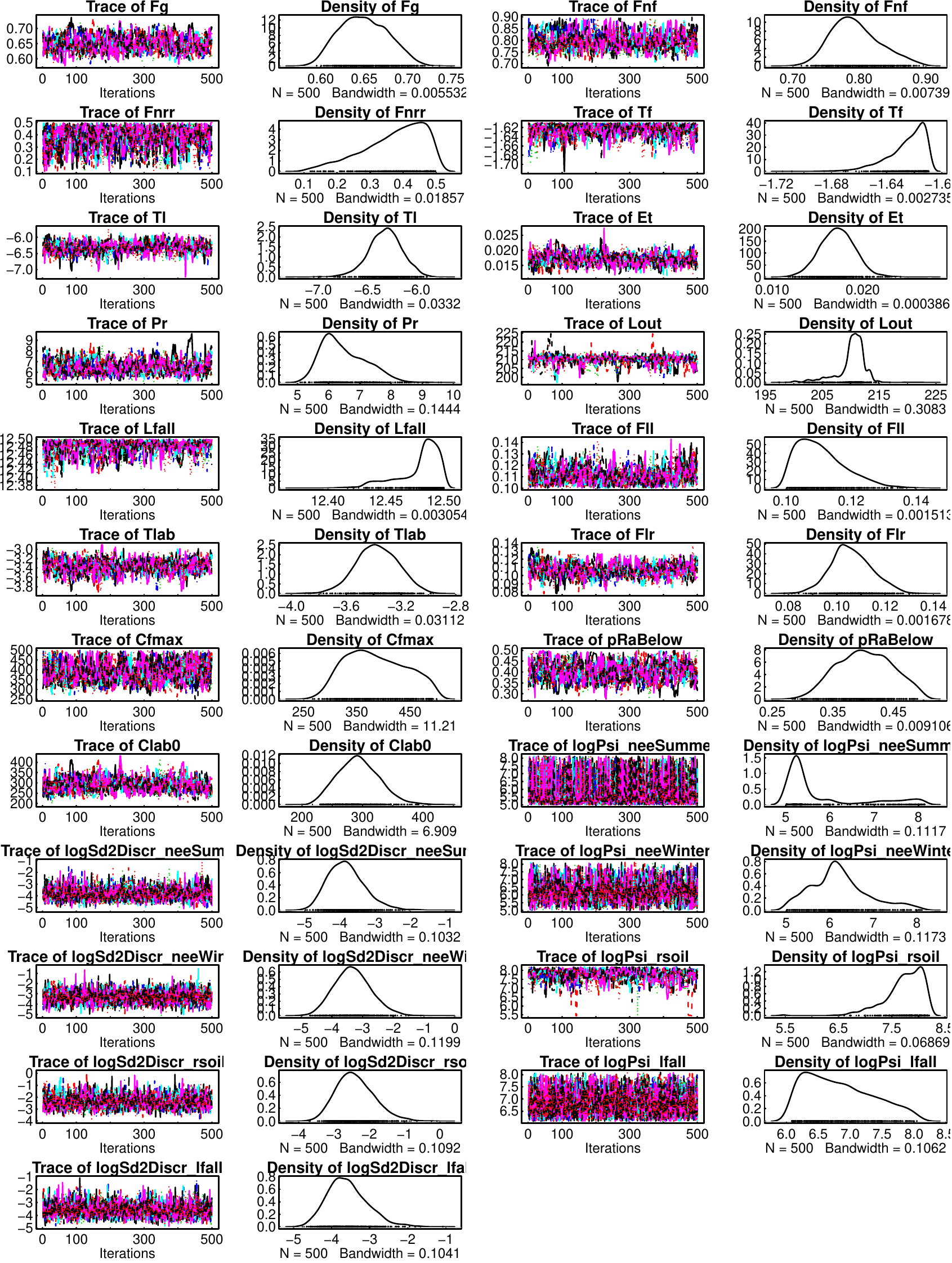}\end{center}
\caption{\label{fig:MCPlotDalec} 
Trace and marginal distribution of the tails of eight Chains from 
two independent populations sampling 15 parameters of the DALEC model. They
indicate good mixing and convergence to the limiting distribution.}
\end{figure}

Eight chains from two independent populations converged to the same limiting
distribution (Figure \ref{fig:MCPlotDalec}).

\begin{figure}
\begin{center}\includegraphics{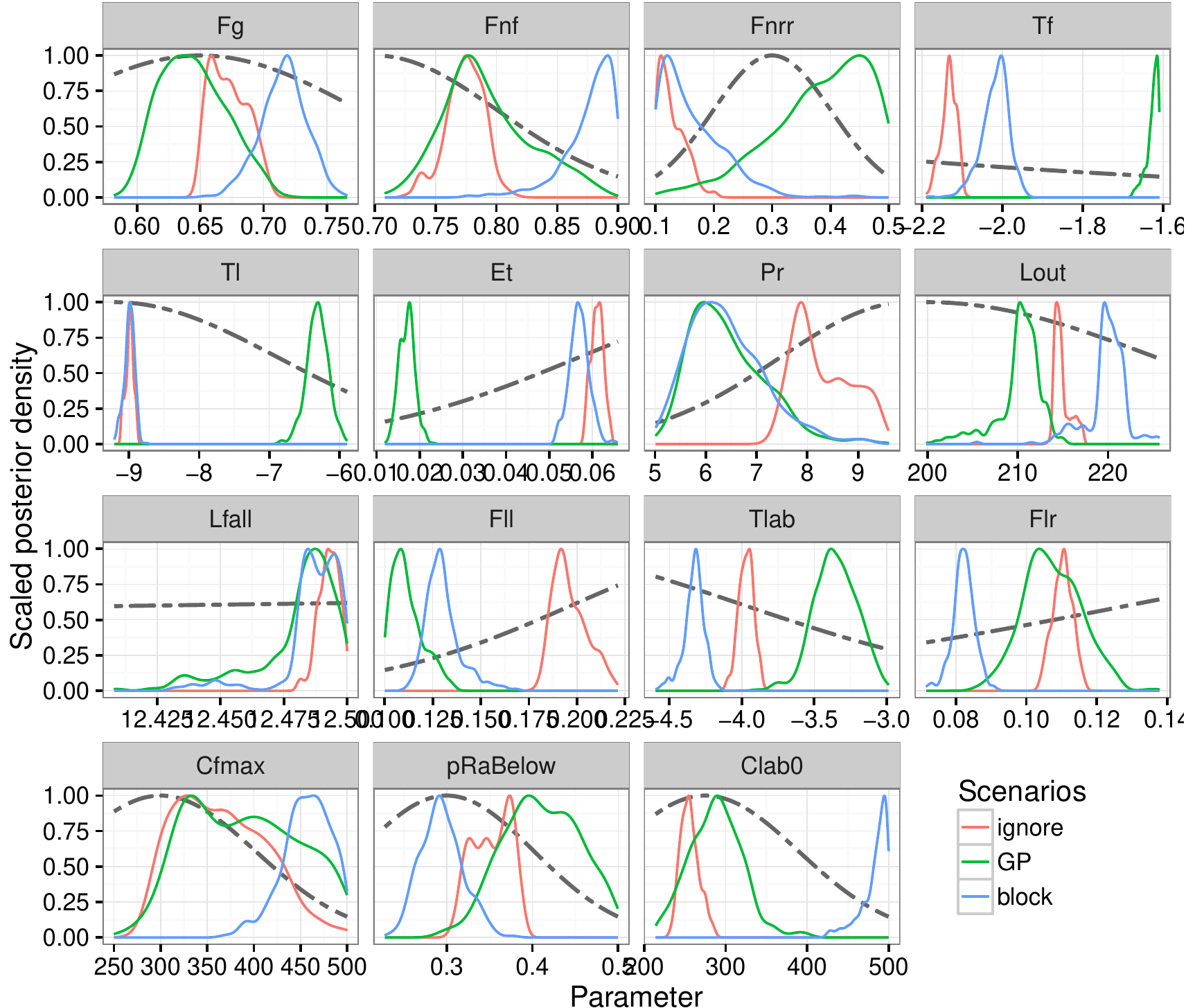}\end{center}
\caption{\label{fig:HowlandPost} 
Marginals of posterior probability density of inverting the DALEC model.
The thick dash-dot line represents prior parameter density. 
}
\end{figure}

The variance of the marginals of posterior probability density inferred by the
GP approach, was intermediate between the ignore and the blocked inversion
scenarios. The location of the mode was mostly in between the two other
approaches (Figure \ref{fig:HowlandPost}).

\begin{figure}
\begin{center}\includegraphics{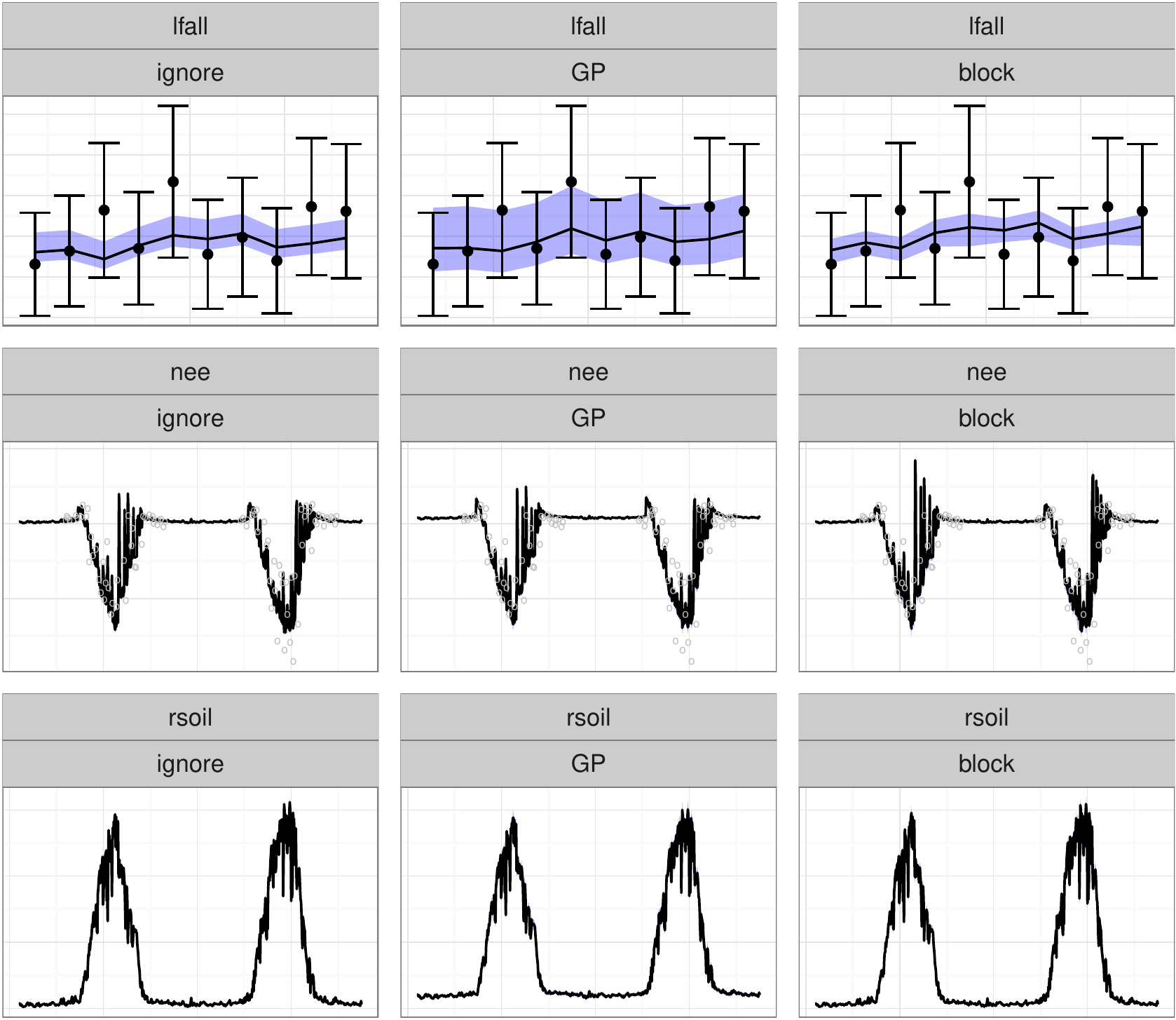}\end{center}
\caption{\label{fig:HowlandPredPost} 
Comparsion of model predictions to observations of the DALEC inversion during
the calibration period. Columns represent different inversion scenarios. ignore:
ignoring model discrepancies, GP: Gaussian processes approach, blocks: parameter
block approach.
Rows show different data streams. Litterfall is shown for predictions of 10
years of calibration period. NEE and respiration are shown for the last two of
the 10 years.
}
\end{figure}

Prediction during calibration period are quite similar across inversion
scenarios (Figure \ref{fig:HowlandPredPost}). Nevertheless there were
differences in uncertainty of posterior predictions
(Figure \ref{fig:HowlandPredPostFuture}).

\ack We thank Nuno Carvalhais and Paul Bodesheim for constructive discussions on
model data integration and on GPs, Adam Erickson for help with the English,
Trevor F. Keenan and David Hollinger for permission to use the Howland data and
model setup.
The work has been partly funded by the Deutsche Forschungsgemeinschaft CRC
1076 "AquaDiva".
\bibliographystyle{dcu}
\bibliography{twutz_txt}  
\end{document}